\documentclass[fleqn,usenatbib]{mnras}

\usepackage{newtxtext,newtxmath}
\usepackage[T1]{fontenc}

\DeclareRobustCommand{\VAN}[3]{#2}
\let\VANthebibliography\thebibliography
\def\thebibliography{\DeclareRobustCommand{\VAN}[3]{##3}\VANthebibliography}

\usepackage{graphicx}	% Including figure files
\usepackage{amsmath}	% Advanced maths commands
\newcommand{\kms} {$\mathrm{ km \; s^{-1}}\,$}
\newcommand{\msol} {M$_{\odot}$}

\newcommand{\mza} {M$_{\rm ZAMS}$}
\newcommand{\zsol} {Z$_{\odot}$}
\newcommand{\rsol} {R$_{\odot}$}
\newcommand{\about} {$\sim$}

\newcommand{\halpha} {$\mathrm{H\alpha}$ }

\newcommand{\hoki}{{\tt hoki}\,}
\newcommand{\mdot}{$\dot{\rm M}$}
\title[Binary pathways to SLSNe-I]{Binary pathways to SLSNe-I: SN 2017gci}

\author[H. F. Stevance]{
H. F. Stevance,$^{1}$\thanks{E-mail: hfstevance@gmail.com}, J. J. Eldridge$^{1}$
\\
$^{1}$The Department of Physics, The University of Auckland, Private Bag 92019, Auckland, New Zealand\\
}

\date{Accepted XXX. Received YYY; in original form ZZZ}

\pubyear{2021}

\begin{document}
\label{firstpage}
\pagerange{\pageref{firstpage}--\pageref{lastpage}}
\maketitle

\begin{abstract}
Some hydrogen poor superluminous supernovae (SLSNe) exhibit bumps in the tails of their light-curves associated with hydrogen features in their late time spectra. 
Here we use the explosion parameters of one such SLSN -- SN~2017gci -- to search the stellar models of the Binary Population And Spectral Synthesis (BPASS) code for potential progenitors. 
We find good matches for a 30\msol\, progenitor star in a binary system and no matches from single star models. % are not able to approach the conditions of the progenitor of SN~2017gci. 
%We also search for and exclude a binary scenario previously described where the late time hydrogen emission is the result of a companion being stripped during the explosion.
Common envelope and mass transfer after the giant branch, combined with increased mass loss from strong stellar winds immediately before death, allow the progenitor to lose its hydrogen envelope decades before the explosion. This results in a hydrogen poor SLSN and allows for delayed interaction of the ejecta with the lost stellar material. 

\end{abstract}


%%%%%%%%%%%%%%%%%%%%%%%%%%%%%%%%%%%%%%%%%%%%%%%%%%

%%%%%%%%%%%%%%%%% BODY OF PAPER %%%%%%%%%%%%%%%%%%

\section{Introduction}
\label{sec:intro}
Superluminous supernovae (SLSNe) are a class of astronomical transients separated from standard supernovae by a luminosity threshold of M$_g<-19.8$ mag \citep{gal-yam19}. Similarly to their less luminous counterparts, SLSNe come in two main flavours: hydrogen rich (SLSNe-II) and hydrogen poor (SLSNe-I). The excess brightness of type II SLSNe is likely powered by interaction of the ejecta with hydrogen rich circumstellar material (CSM -- see \citealt{gal-yam19} for more details), whereas the case of the SLSNe-I still remains under debate.
The most commonly invoked mechanism is the magnetar model in which the light curve is powered by the spin-down of a strongly magnetic (B\about10$^{14-15}$G), rapidly rotating (P\about1-4ms) central engine \citep{woosley10, kasen10, nicholl17}. This model has been largely successful at reproducing the early lightcurves of hydrogen poor SLSNe, particularly those with smooth decay tails (e.g. \citealt{dessart12, inserra13, nicholl15}). 

There is, however, a number of hydrogen poor SLSNe which show oscillations in their late time lightcurve (a few tens to over 100 days after maximum, e.g iPTF13ehe, SN2015bn, iPTF15esb, iPTF16bad, SN2017gci-- \citealt{yan15, nicholl16, yan17, fiore21}), some of which accompanied by detections of \halpha in their spectra . 
The presence of these features could be a sign of interaction with CSM \citep{yan17, inserra17}, e.g. from pulsational pair instability a few decades before explosion (iPTF13ehe,  \citealt{yan15} -- iPTF16ehe, \citealt{lunnan18}).
Alternatively \cite{moriya15} developed a binary model whereby the broad hydrogen features (\about 4500\kms in \citealt{yan15}) are the result of mass stripping from a binary companion by the supernova ejecta. The hydrogen material is initially hidden within the photosphere, but as the opacity decreases, the hydrogen is revealed during the nebular phase. 

In this work we focus on the recently published SN 2017gci \citep{fiore21}, which showed two rebrightenings at \about103 and \about142 days after maximum light and the emergence of \halpha from +51 days to \about133 days.
\cite{fiore21} performed fits on the photometric data using three models: a nickel powered, magnetar powered and CSM powered lightcurve. 
The nickel model was excluded as it required a larger nickel mass than the derived ejecta mass for SN~2017gci; a CSM model with $M_{\rm ej} \approx$ 12 \msol\ and $M_{\rm csm} \approx$ 5 \msol\ can reproduce the peak, but would require an additional power source to explain the tail; a magnetar model with initial period P$\approx$2.8ms, and ejecta mass $M_{\rm ej} \approx$ 9-10 \msol, on the other hand, provided a good fit to the peak and the tail.
We seek to find models that can give rise to the progenitor and environment of SN~2017gci. 
Here we focus on the scenario where the bumps and associated hydrogen features originate from CSM around the progenitor \citep{fiore21}.

We use the reported parameters of the explosion to create a set of criteria and search existing stellar models in the Binary Population and Spectral Synthesis (BPASS -- \citealt{eldridge08, eldridge17, stanway18}) code.
These simulations have a wide scope and have been used in a large array of applications: to better understand and age stellar populations (e.g. \citealt{wofford16, stevance20, brennan21}), study nebular emission (e.g \citealt{xiao18, xiao19}), investigate the reionization of the universe (e.g \citealt{stanway16, ma16}), and predict binary black-hole mergers \citep{eldridge16}, to name a few.
Our main goal is to check whether matching progenitors can be found natively in stellar evolution models that self-consistently recreate the characteristics of stellar populations found in the Universe.

In Section~\ref{sec:data} we give a brief description of the data used in this work, in Section~\ref{sec:search} we lay out a set of search criteria based on the explosion parameters of SN~2017gci and the literature. In Section~\ref{sec:results} we describe the matching models, these results are discussed in Section~\ref{sec:disc} before concluding and summarising our findings in Section~\ref{sec:conc}.

%The latter are sometimes subdivided into a rapidly-declining and a slow-declining category \citep{gal-yam19} although a clear separation between these classes is not consistently observed \citep{nicholl15, nicholl17, decia18}.

\section{Model Data}
\label{sec:data}
We use the stellar models of BPASSv2.2.1\footnote{\url{https://drive.google.com/drive/folders/1BS2w9hpdaJeul6-YtZum--F4gxWIPYXl}}
which  are computed for 13 metallicities and 9 different initial mass functions (IMF -- for more details see \citealt{eldridge17, stanway18}).
Here we only consider the fiducial IMF: a Kroupa prescription with maximum mass 300 \msol\, \citep{kroupa01}.
BPASS creates a population of stars amounting to 10$^6$\msol\, according to the chosen IMF and evolves them over 51 time bins ranging form log(age/years)=6.0 to log(age/years)=11.0 in increments of 0.1 dex. 
This includes single-star-only models as well as binary models; here we focus on the latter. 
It is important to note that these binary models also contain single stars (and effectively single stars) that do not undergo binary interaction, as is expected in the Universe.
Additionally, the distribution of initial periods and mass ratios are based on observations \citep{moe17}.
In the following searches we will not explicitly select for binary or single system.

BPASS currently includes \about250,000 stellar models, each recorded in one text file compiling over a hundred properties over dozens to hundreds of time steps.
With modern machines, computational bottlenecks often reside in memory and disk access rather than the number of floating point operations that processors can handle. 
In order to avoid repeatedly opening and reading thousands of text files we used \hoki \citep{hoki} to compile the stellar models into a data product better suited to data analysis within python. 
Specifically the data were split into 26 binary files each containing one {\tt pandas} dataframe \citep{reback2020pandas, mckinney-proc-scipy-2010} comprised of all the stellar models for one metallicity and one simulation type -- either single star models only, or including binaries \citep{stevance20_df}\footnote{https://zenodo.org/record/4064300\#.YDyH5nVfgYt}. 
These data products can then be searched quickly and efficiently using existing python infrastructure and they are made widely accessible so that similar analyses can be performed by others in the future.

%\subsection{BPASS }

%\subsection{{\tt hoki} code and data products}

\section{Search criteria}
\label{sec:search}
\begin{table}

 \caption{\label{tab:summary_constraints}Summary of the model search constraints and the number of stellar models left after each is applied. In the third column the number of models are given for an explosion energy 10$^{52}$ ergs (10$^{51}$ ergs).}
 \label{tab:anysymbols}
 \begin{tabular*}{\columnwidth}{l l l }
  \hline
  Property & Constraint & Number of models\\
  \hline
  Z & 0.006 & 21,598\\
  M$_{\rm ZAMS}$ & $>$10\msol & 9476\\
  M$_{\rm rem}$ & $<$2.43\msol &  9233 (5155) \\
  M$_{\rm ej}$  & 8--13\msol & 1863 (774)\\
  X (at death) & $<$0.01 & --\\
  M$_H$ (at death) & $<$0.001 \msol & 761 (2)\\
  %j$_*$ (at death) & $>$ 10$^{14}$ cm$^2$.s$^{-1}$ & 651 (1)\\
  t$_{\rm csm}$ & $<$ 150 years before death & 2 (None)\\
  %M$_He$ (at death) & $<$0.32 \msol & -- --\\
  \hline
 \end{tabular*}
\end{table}

\subsection{Basic constraints}
First we restrain our search to a single metallicity since \cite{fiore21} estimate that the host galaxy of SN 2017gci has Z $\sim$0.3\zsol\,based on strong line indicators (D02, PP04 O3O2, PP04 N2Ha,M13N2 \citealt{denicolo02,pettini04,maiolino08, marino13}). Assuming a solar metallicity mass fraction of 0.02, the closest match in BPASS are the Z=0.006 models.

Next, we apply a \mza\, mass cut-off. \cite{fiore21} state that the progenitor is likely over 40 \msol\, but this is based on the single star models  of \cite{jerkstrand17}, and so as not to bias this search a lower mass constraint of 10 \msol\,is chosen.
In theory no mass cut-off need to be applied, but in practice some of the following steps are unnecessarily computationally intensive without this limit. This makes the data more manageable without introducing a bias since the ejecta mass in \cite{fiore21} is \about 10\msol\, and our limit is therefore very conservative.

\subsection{Ejecta and remnant mass}
Next the ejecta mass and the remnant mass are constrained.
The former is narrowed down by \cite{fiore21}  to $M_{\rm ej}\approx$9\msol\, or $M_{\rm ej}\approx$12\msol\, for the magnetar and CSM models respectively. 
The former is preferred as it can explain the peak and tail of the light-curve of SN~2017gci.
Not knowing the full extent of the uncertainties on these ejecta masses we keep our search wide and  constrain our $M_{\rm ej}$ to be between 8\msol and 13\msol.

%In this work neither model is preferred and we consider a case where a mixture of both models is required to understand the explosion properties of SN 2017gci.As a result the ejecta mass constraint encompasses both estimate, requiring a $M_{\rm ej}$ between 8\msol and 13\msol. 

Since a magnetar is also required, the remnant mass is limited by the maximum mass of a neutron star (NS). This upper limit remains uncertain but studies based on the analysis of the NS-NS merger GW170817 \citep{abbott17, abbott19} report a 90 percent upper bound at 2.43 \msol\citep{abbott18}; other works by e.g. \cite{essick20} find a slightly lower mass, around 2.3 \msol (for an extensive discussion of the uncertainty on the NS mass limit see section 6.2 of \citealt{abbott20}).
For the purposes of this search 2.43 \msol\, was used as the limiting value, but given the remnant masses of the model matches (\about 1.4\msol, see Section \ref{sec:results}), the uncertainty in the NS maximum mass is not a cause of concern. 

It is important to remark that the remnant mass and ejecta mass in BPASS (and in the Universe) are dependent on the energy output of the explosion.
In our models there are three sets of ejecta and remnants masses corresponding to ``weak" (10$^{50}$ ergs), ``normal" (10$^{51}$ ergs) and ``super" (10$^{52}$ ergs) supernovae explosions -- note these keywords are used in the BPASS manual and in the \hoki data products. 
\cite{nicholl17}, for a sample of 38 hdyrogen poor SLSNe, found kinetic energies ranging from 2$\times$ 10$^{51}$ to 10$^{52}$, so in the rest of this study we ignore the ``weak" explosions.

\subsection{Hydrogen content at death}
%A final key characteristic that must be reproduced by the models is the hydrogen content of the progenitor of SN 2017gci. 
Because SN 2017gci is a SLSN-I, the hydrogen levels of its progenitor at death must be close to none.
Specifically, simulations by \cite{dessart12} showed that strong \halpha lines can be present in supernovae spectra with a hydrogen mass fraction (X) of 0.01 and a total hydrogen mass (M$_{H}$) of 0.001\msol. These values are used here as boundaries between  hydrogen ``rich" and hydrogen ``poor" progenitors. 

\subsection{Hydrogen content before death}
\label{sec:H_content}

The presence of bumps in the lightcurve of SN~2017gci as well as \halpha signatures in its spectra between phase 51 and 133 days are interpreted as the presence of hydrogen rich CSM, potentially in shells \citep{fiore21}. 
If this CSM originates from the progenitor of SN 2017gci, we can roughly estimate when the final layers of hydrogen must have been lost by following
\begin{equation}
    {\rm t}_{\rm csm} = \frac{{\rm v}_{\rm phot}\times {\rm t}_{\rm interact}}{{\rm v}_{\rm csm}},
\end{equation}

where t$_{\rm csm}$ is how long before the explosion the CSM was ejected, v$_{\rm phot}$ is the photospheric velocity (estimated by \cite{fiore21} to be \about 8000\kms), t$_{\rm interact}$ is the time with respect to explosion when the ejecta encounters the CSM, and  v$_{\rm csm}$ is the velocity of the CSM.

Not all of these quantities are known, specifically the CSM velocity is not quantified in \cite{fiore21}, and the explosion date is uncertain. 
Based on SLSN-I samples from the PAN-STARRS1 Survey and the Palomar Transient Factory, the rise time of this type of explosion ranges from as low as 15 days to \about 100 days \citep{decia18, lunnan18}. 
In the case of SN 2017gci, the rising phase of the lightcurve is not captured; for our estimates we focus on two cases, t$_{\rm rise}\sim50$ days  and t$_{\rm  rise}\sim100$ days, and since the first sign of \halpha is seen at 51 days after maximum, these correspond to t$_{\rm interact}$ of 100 and 150 days.
Finally, for the CSM velocity, a slow and a fast wind scenario are considered:  v$_{\rm csm}=100$ \kms and  v$_{\rm csm}=1000$ \kms. 
Wind velocities of heavily stripped WR stars can easily surpass the velocity of the fast wind case (e.g. 5000\kms -- see \citealt{tramper15} and references therein), but here we are interested in an order-of-magnitude estimate. 

For t$_{\rm interact}$=100 days the final layers of hydrogen would have been expelled \about 20 years (\about2 years) before explosion given a CSM velocity of v$_{\rm csm}=100$ (1000)\kms.
In the case of a longer rise time and t$_{\rm interact}$=150 days, these values rise to \about 30 years (\about3 years). 
Overall we are looking for progenitors that would have lost their final hydrogen layers a matter of decades at most before the explosion. 
To keep our search broad, since many WR stars have wind velocities of several thousand \kms, we select all models which lost the last of their hydrogen 150 years before death.

\iffalse
\subsection{Additional constraints if H came from a companion}
The model of \cite{moriya15} places limits on the separation of the system and the mass of the secondary star.
They estimate the companion stripping fraction for separations ranging from 5 to 10R$_2$ and note that for separations of 20R$_2$ or above the companion mass needs to be greater than 1000 \msol. This is an extreme limit that would be unphysical for SN~2017gci so we constrain the system separation at death to be $\le$ 10R$_2$.
As for the mass of the secondary, \cite{moriya15} find different minimum mass requirements to match iPTF 13ehe for different explosion mechanism: 20 \msol for a nickel powered model and 80 \msol for a magnetar spin-down model. 
It has been determined by \cite{fiore21} that nickel is not a viable energy source for SN~2017gci, but we nevertheless use the $M_2\ge$20\msol mass limit at this stage to allow for further discussion in Section \ref{sec:h_from_comp}.
\fi 

\section{Results}
\label{sec:results}
%\subsection{H from the progenitor}
After performing model selection according to the criteria described in the previous section, we find 2 excellent matches: a 30 \msol\ primary with a 12\msol\ binary companion and a 30\msol\ secondary star with a 25\msol\ black-hole (BH) -- see Table \ref{tab:summary_properties} for a full summary of their properties. Both models follow a very similar evolutionary pathway; the main difference resides in the nature of the companion  but qualitatively their life is identical and quantitatively they differ only marginally. The scenario with the main sequence companion is much more likely to occur from an IMF stand-point -- 25\msol\ black holes are rarer than 12 solar mass MS stars. We can go further and quantify the odds of the primary model occurring over the secondary: one of the properties given for each BPASS stellar models is the number of stars that would be found in a 10$^6$\msol\ population with the characteristics of that specific model. 
We find that the primary (secondary) progenitor model would apply to 1.58 (0.014) stars per 10$^6$\msol\ , meaning that the odds of the first scenario occurring over the latter are \about 110:1. 
Consequently, in the rest of this section, we focus on the primary model.

\begin{figure}
\centering
 \includegraphics[width=6.5cm]{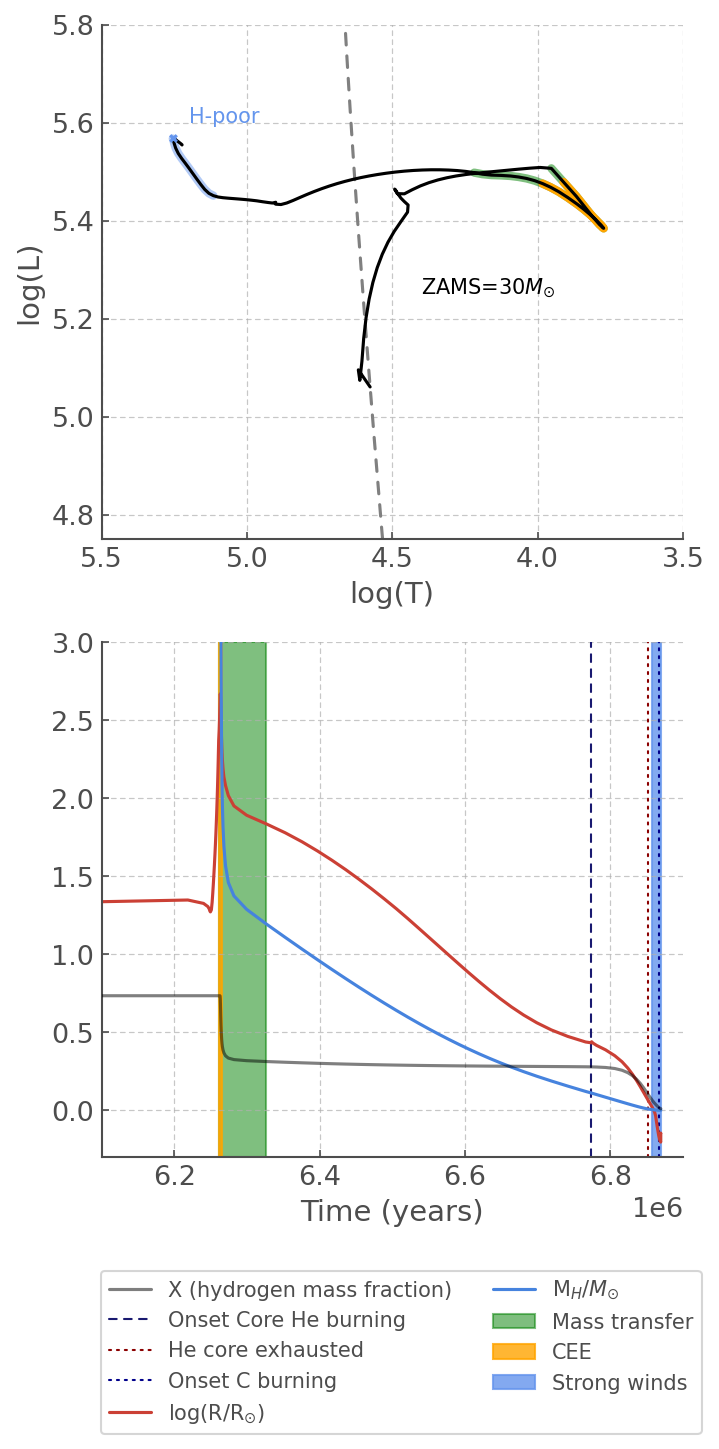}
 \caption{\textit{Top panel:} Evolutionary tracks of the matching model. \textit{Bottom panel:} 
 evolution of the hydrogen content and radius from the end of the main sequence to the death of the star.}
 \label{fig:evolution}
\end{figure}

Figure \ref{fig:evolution} shows the evolutionary track of the 30\msol\ star  as well as its hydrogen content and radius evolution from the end of the main sequence onward. 
The system begins its life in a \about5-month orbit with a 428 \rsol\, separation.
The main sequence is uneventful and the two stars essentially follow single star evolution, but after the onset of hydrogen shell burning, the radius of the primary increases rapidly up to 468 \rsol\,  (see Figure \ref{fig:evolution}). The transition to the red giant branch begins a common envelope phase that lasts 3800 years and strips the primary star of 12\msol\, of material. 
This mass loss episode changes the chemical make up of the progenitor -- nearly 8 \msol\, of hydrogen are lost to the interstellar medium and the hydrogen mass fraction falls from 0.74 to 0.43. 
As a result of the mass loss, the radius of the progenitor stars decreases and the common envelope phase ends once it drops below \about 250 solar radii. 
Roche lobe overflow and mass transfer continue for a further 60,000 years, causing an additional loss of 1.8 \msol\,  from the stellar envelope (0.75\msol\, of which hydrogen). The hydrogen mass fraction decreases marginally during this phase by $<$0.1. 

When the system detaches its period is \about 50 days and the progenitor has lost over half of its \mza, but the hydrogen content is still insufficiently low to explain a type I SN (or SLSN) explosion. Mass loss continues in steady winds with \mdot=5.95($\pm$ 1.4) $\times 10^{-6}$ \msol/yr  for the next 0.5 Myrs, but a dramatic increase in luminosity in the last 6000 years raises \mdot\, to 2 $\times 10^{-5}$ \msol/yr.
This rise in luminosity begins after the exhaustion of the helium core and the onset of Helium shell burning, and it continues through carbon core burning (see Figure \ref{fig:evolution}).
It is this late rapid mass loss event that allows the hydrogen mass to drop below the threshold of 0.001 \msol, which is reached 54 years before explosion.
With a kinetic energy of 10$^{52}$ the ejecta mass reaches 10\msol which matches the ejecta mass of SN~2017gci very well despite our loose constraints.

The helium content at the end of our simulations amounts to 1.54\msol\, or 13.4 percent of the total mass of the model. 
It is worth noting that the BPASS models do not compute the full nucleosynthesis all the way to iron (in this model steps beyond the start of neon burning 3.5 months before explosion are not captured by the models) therefore the final helium content at explosion could differ slightly. 
On the whole this echoes \cite{yan17} who remarked that if the hydrogen must be lost decades before the event, the progenitor will not have time to lose its entire helium envelope and it is likely that the ejecta of such SLSNe-I do contain helium. 
Large helium masses could remain invisible if there is not enough nickel to provide the non-thermal excitation  required to excite helium lines \citep{lucy91} .
In the context of type Ib/c SNe, \cite{dessart12} showed that even different levels of mixing could be enough to hide over 1\msol\,of helium in an ejecta 4-5\msol. 
Therefore it is not unreasonable that 1.5\msol\,of helium would go unnoticed in a magnetar dominated SLSN with 10 \msol\, of ejecta. 

%\subsection{The secondary model}

%30 \msol progenitor with a 25 \msol BH in a $\apporx$ 2-month orbit separated by 254\rsol. The evolution is very similar to the primary model. At its largest is 363 \msol - the CEE lasts 2775 years. During CEE 10 \msol are lost, 6.7 of which are H, and X drops by 0.25.in the rest of the mass transfer which lasts 65 k years, another 2.6 \msol of material are lost of which 1.1 are H, and the X fraction drops by 0.14 to 0.31. then relatively smooth mass loss in steady winds with \mdot=5.90(\pm 1.2)$\times 10^{-6}$\msol/yr  nut the ;ast leg of h loss happens as L dramatically increases as m dot rises to 2$\times 10^{-5}$\msol/yr just after the onset of carbon core bruning. this rapid mass loss highlighted is where mdot >1e-5 in the last 6300 years and that's what allows Mh1 too drop below our threshold

\begin{table}
 \caption{\label{tab:summary_properties}Properties of the two matching models.}
 \label{tab:anysymbols}
 \begin{tabular*}{\columnwidth}{l l l }%{@{}l@{\hspace*{50pt}}l@{\hspace*{50pt}}l@{}l@l@l@}
  \hline
  Property & Primary progenitor & Secondary progenitor \\
  \hline
  
  Primary M$_{\rm ZAMS}$  & 30 \msol & 25 \msol\, BH\\
  Secondary M$_{\rm ZAMS}$  & 12 \msol& MS star 30 \msol\\
  P$_{\rm ZAMS}$  & 158 days & 63 days \\
  M$_{\rm ej}$  & 9.97 \msol& 10.0 \msol\\ 
  M$_{\rm rem}$  & 1.45 \msol& 1.46 \msol\\ 
  X   & 0.0098 & 0.0099 \\ 
  M$_H$  & 0.0004 \msol& 0.0004 \msol\\
  t$_{\rm csm}$  & 54 years & 31 years \\
  P$_{\rm death}$  & 64 days & 57 days\\
  %j$_*$ & 2.80 $\times$ 10$^{15}$ cm$^2$.s$^{-1}$ & 3.12 $\times$ 10$^{15}$ cm$^2$.s$^{-1}$\\
  \hline
 \end{tabular*}
\end{table}

\iffalse
\subsection{H from the companion}
\label{sec:h_from_comp}
With the secondary mass limit of 20\msol we find 59 matching models, however this is based on a nickel-powered SLSN explosion. 
In the case of magnetar powered explosion for iPTF13ehe \cite{moriya15} find that a companion of more than \about 80 \msol is necessary. 
In our 59 models, the maximum secondary mass at death is 62\msol.
Consequently, the scenario in which the hydrogen in the late time light curve of SN~2017gci originates from a binary companion being stripped during the explosion such as described in \cite{moriya15}. 
%It is possible that alterations to that scenario may allow for better fits to the observations but it is beyond the scope of this paper to explore this. 
\fi

\section{Discussion}
\label{sec:disc}
We used the explosion parameters derived by \cite{fiore21} for SN~2017gci to define search criteria to look for potential progenitors in the BPASS models (including both single and binary systems). 
The two matching stellar models we found were binary systems and if we perform the search on single-star-only models, no matches can be found. 
One of the crucial differences is in the nature of the remnant: where the binary models selected for NS remnants return 5851 (1861) matches for an explosion energy of 10$^{52}$ (10$^{51}$) ergs, the single star models only provide 84 (3) matches. None of these are hydrogen free at the end of their lives. 
Tidal stripping in binary interactions is therefore an essential mass-loss step to yield the matching models. 
This is in line with the known importance of binary interactions for the production of other types of stripped supernovae \citep{dessart12, yoon15}.

It is also interesting that the ``normal" supernova (10$^{51}$ ergs) prescription rapidly chokes the number of matching models (see Table \ref{tab:summary_constraints}). That is expected since a high kinetic energy input in the progenitor results in a lower remnant mass (making it easier to achieve the NS requirement) and a higher ejecta mass. 
Overall, this agrees with the kinetic energy of observed SLSNe (e.g. \citealt{nicholl17}) which is typically greater than 10$^{51}$ ergs.

One of the main unknowns in our models  is the precise description of the CSM resulting from the strong winds at the end of the life of the progenitor. Ideally the environment must be capable of producing the spectral lines and rebrightnening observed in SN~2017gci; CSM mass, opacity and geometry will all play a role. 
Modelling and deriving these parameters would be helpful in general, but could not aid in defining further search criteria in the BPASS models since the wind prescriptions used are not sufficiently detailed  for this type of work.
One of the main questions is that of the origins of the multiple brightening events observed in this type of SLSNe (e.g iPTF13ehe, SN2015bn, iPTF15esb, iPTF16bad -- \citealt{yan15, nicholl16, yan17}).
If they are indeed the result of shells of CSM, as has been previously suggested \citep{yan15, yan17}, then the stellar wind scenario needs to be able to produce shell-like geometries. 
Bow-shocks and photo-ionization confined shells could provide a natural explanation for this \citep{wilkin96, mackey14} but it would be strongly dependent on the environment and/or speed of the progenitor. 

The potential presence of CSM around SLSN progenitors has previously been attributed to the pair-instability (PI) mechanism (for details on the mechanism see \citealt{woosley17}).
This is an attractive scenario as the CSM and bright explosion are both a consequence of the same phenomenon.
However, the CSM and ejecta masses resulting from PI are much higher (e.g. \citealt{yan15}) than observed in the case of SN~2017gci and most PI explosions are nickel powered (which was excluded by \citealt{fiore21}) or CSM powered (which did not match the tail of the lightcurve).
Our models can therefore fill areas of parameter space not covered by the PI models and these scenarios do remain relevant in a higher mass regime.

Finally, hydrogen poor SLSNe with bumpy tails are quite common and we therefore need any candidate pathway to be able to reproduce the numbers observed. Doing this in detail is not possible at this stage because, as mentioned above, it is likely that more than one pathway contributes to this type of transient, and it would require careful consideration of the star formation history associated with their host galaxy. 
That being said we can look at how rare the best matching model is using the model IMF calculated by BPASS: that is the number of systems one expects to find per million solar masses. 
For our primary model we find a model IMF of 1.58, so for a host galaxy of mass between 10$^8$ and 10$^9$ \citep{perley16}, we expect between 150 and 1500 such systems if we assume a single burst of star formation.
Consequently, the pathway described here is not rare by any means.

\section{Conclusion}
\label{sec:conc}
In this study we created search criteria (see Table \ref{tab:summary_constraints}) based on the explosion parameters of the hydrogen-poor superluminous SN 2017gci derived by \cite{fiore21} to find potential progenitors in the BPASS models. 
We found two \mza=30\msol stellar models matching our constraints. 
Both are found in binary systems and single star models were not able to meet the constraints laid out in Section \ref{sec:search}.

\begin{figure}
 \includegraphics[width=\columnwidth]{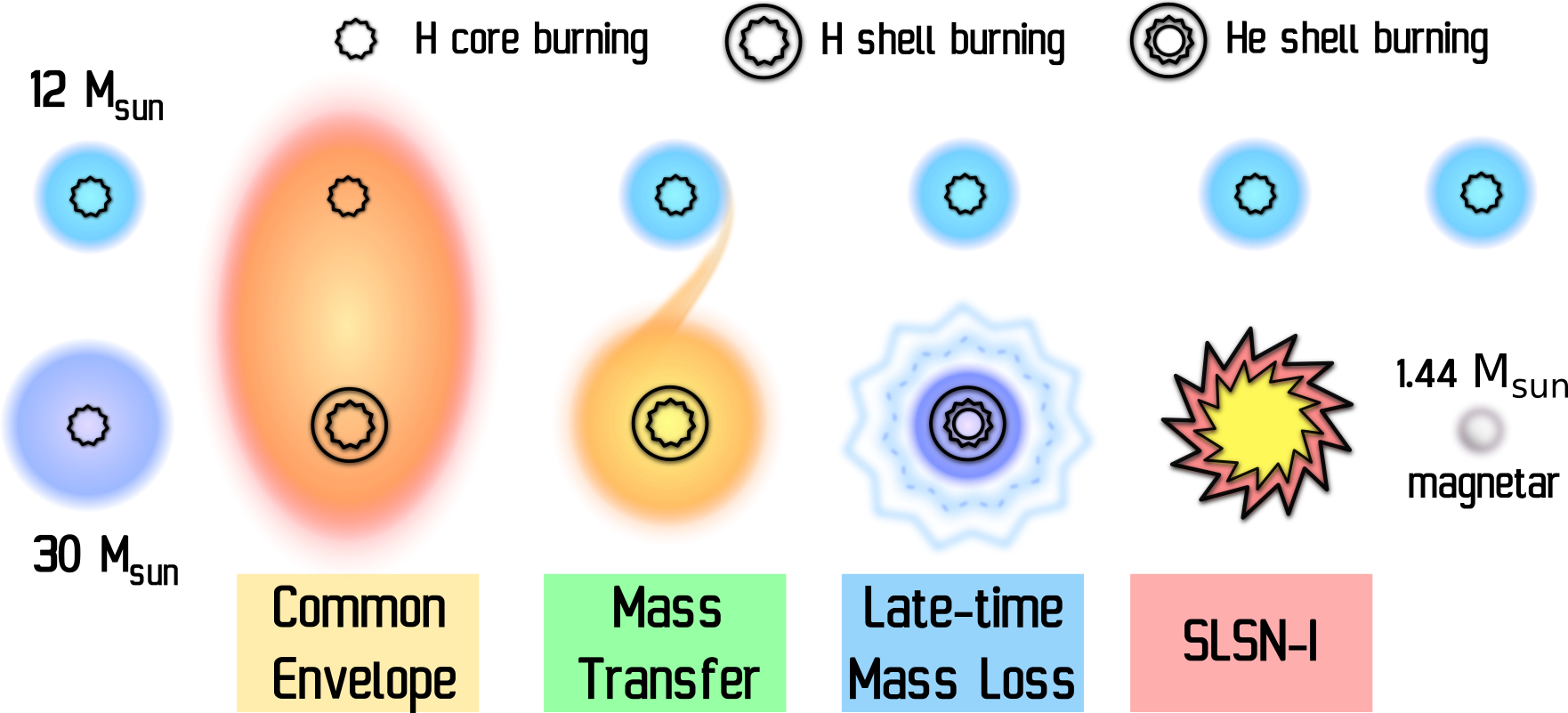}
 \caption{Evolutionary Pathway to the hydrogen poor superluminous SN~2017gci}
 \label{fig:cartoon}
\end{figure}

The most likely model (see Figure \ref{fig:cartoon}) has a 12\msol main sequence companion, whereas the other is a secondary star in orbit with a 25 \msol black hole primary. We find that both progenitors have very similar pathways with a phase of common envelope evolution starting when the progenitor reaches a size of \about 250\rsol\, followed by further mass transfer. Mass loss continues in winds and increases dramatically shortly before death as a result of a drastic increase in luminosity.

These models are interesting for events such as SN~2017gci where a nickel powered explosion is unphysical and the ejecta mass relatively low (\about 10 \msol).
This renders scenarios such as pair-instability CSM previously invoked to explain the bumpy light-curve and late hydrogen features \citep{yan15} inapplicable.
It is also interesting to note that the progenitor here has a relatively low mass compared to other progenitor studies (e.g. \citealt{jerkstrand17}). 
The model in which hydrogen originates from a stripped companion described in \citep{moriya15} used ejecta masses and energies very different from those found in SN~2017gci (see their table 1). A direct comparison is not appropriate as it was devised to match the specific observable of iPTF13ehe, so we cannot comment on the applicably or inapplicability of this scenario to SN~2017gci specifically, but it is worth noting that significant companion stripping from shock heating is expected for close companions (e.g. \citealt{hirai14, hirai18}) and the \citeauthor{moriya15} model could have larger applications. 

A crucial difference between our progenitor search and that of other studies is that we do not explore an unconstrained parameter space to create a system that will best match a particular event; instead we searched existing simulations that simulate  realistic stellar populations.
Dedicated models can test specific aspects of the physics with more accuracy whilst wide scope simulations such as BPASS can check that certain pathways are consistent with stellar evolution at large. Both approaches are therefore essential if we are to fully uncover the progenitors of SLSNe.

\iffalse

\fi

\section*{Acknowledgements}
The authors would like to offer warm thanks to the referee for their careful and insightful review.
HFS would like to thank S.G. Parsons for interesting conversations and for their enduring support. HFS and JJE acknowledge the support of the Marsden Fund Council managed through Royal Society Te Aparangi. We are grateful to the developers of matplotlib, numpy \citep{matplotlib, numpy} and pandas.

%%%%%%%%%%%%%%%%%%%%%%%%%%%%%%%%%%%%%%%%%%%%%%%%%%
\section*{Data Availability}
A Jupyter notebook containing our analysis is available\footnote{\url{https://github.com/UoA-Stars-And-Supernovae/Binary_pathways_to_SLSNe_I_17gci}}.
The core data used for this work is contained in the following files in the data directory \cite{stevance20_df}\footnote{https://zenodo.org/record/4064300\#.YDyH5nVfgYt}: {\tt all\_z\_bin\_bin\_models}, {\tt all\_z\_bin\_sin\_models}



\bibliographystyle{mnras}
\bibliography{slsn} % if your bibtex file is called example.bib

%\begin{thebibliography}{99}
%\end{thebibliography}

%%%%%%%%%%%%%%%%%%%%%%%%%%%%%%%%%%%%%%%%%%%%%%%%%%

%%%%%%%%%%%%%%%%% APPENDICES %%%%%%%%%%%%%%%%%%%%%

%\appendix

%%%%%%%%%%%%%%%%%%%%%%%%%%%%%%%%%%%%%%%%%%%%%%%%%%

% Don't change these lines
\bsp	% typesetting comment
\label{lastpage}
\end{document}